\documentclass[12pt,preprint]{aastex} 
 
\title{A large stellar evolution database for population
synthesis studies. V. Stellar models and isochrones with 
CNONa abundance anticorrelations}
 
\author{Adriano Pietrinferni, Santi Cassisi}
\affil{Osservatorio Astronomico di Teramo, Via M. Maggini,
64100 Teramo, Italy; pietrinferni, cassisi@oa-teramo.inaf.it}

\and

\author{Maurizio Salaris, Susan Percival}
\affil{Astrophysics Research Institute, Liverpool John Moores University,
Twelve Quays House, Birkenhead, CH41 1LD, UK; ms, smp@astro.livjm.ac.uk}

\and

\author{Jason W. Ferguson}
\affil{Physics Department, Wichita State University, Wichita, KS 67260-0032, USA - jason.ferguson@wichita.edu}
 
\begin{document}

\begin{abstract}
 
\noindent
We present a new grid of stellar models and isochrones for old stellar populations, covering a large range of [Fe/H] values, for an heavy element 
mixture characterized by CNONa abundance anticorrelations as observed in Galactic globular cluster stars.
The effect of this metal abundance pattern on the evolutionary properties of low mass stars, from the main sequence to the horizontal 
branch phase is analyzed. 
We perform comparisons between these new models, and our reference $\alpha-$enhanced calculations, and discuss 
briefly implications for CMDs showing multiple main sequence or subgiant branches.
A brief qualitative discussion of the effect of CN abundances on color-$T_{eff}$ transformations is also presented, highlighting the need 
to determine theoretical color transformations for the appropriate metal mixture, if one wants to interpret observations in the 
Str\"omgren system, or broadband filters blueward of the Johnson $V$-band.
\end{abstract}
 

\section{Introduction}

\noindent

Libraries of stellar models and isochrones covering wide ranges
of age and initial chemical composition are an essential tool to investigate the
properties of resolved and unresolved stellar populations. 
This is the fifth paper in a series devoted to provide the astrophysics community with an extended, complete, 
and up-to-date database of theoretical stellar models and related byproducts such as isochrones, 
luminosity functions (LF), synthetic Color-Magnitude Diagrams (CMDs), integrated spectra and magnitudes. 

In the first paper of this series (Pietrinferni et al.~2004, hereafter
Paper I) we have presented our database (BaSTI, a Bag of Stellar Tracks and Isochrones) of scaled-solar 
stellar evolution models and isochrones. Later on we have provided (Pietrinferni et al.~2006, hereafter
Paper II) models for an $\alpha$-enhanced metal mixture typical of Galactic Population~II stars 
(see, e.g. Ryan, Norris \& Bessel~1991; Carney~1996; Gratton, Sneden \& Carretta~2004). 
Online tools to apply the model database for studying stellar populations 
(isochrone, LF, and synthetic CMD generator) as well as the extension of the scaled-solar and $\alpha$-enhanced models to the full thermal pulsing 
Asymptotic Giant Branch phase, have been presented in Cordier et al.~(2007, hereinafter Paper III).  
Theoretical integrated spectra and magnitudes of simple stellar populations, based on the BaSTI model database 
have been calculated more recently by Percival et al. (2009, hereinafter Paper IV). 

A further extension of our theoretical scenario to cover an additional metal mixture appropriate for astronomical objects, is presented in this work, where 
we describe models and isochrones computed for an heavy element distribution that includes the CNONa abundance anticorrelations observed 
in Galactic globular cluster (GGC) stars. 

Several decades have passed since the first identifications of star-to-star abundance variations of light elements in stars within 
individual GGCs (Osborn 1971; Norris \& Freeman 1979). Hydrogen burning via the CNO, Ne-Na, and Mg-Al chains at high 
temperatures qualitatively accounts for the observed trends: the abundances of C and O are low when N is high; O and Na are also anticorrelated, 
as are Mg and Al (see Salaris, Cassisi \& Weiss 2002, for a review on this topic). While the amplitude of the anticorrelations may differ from 
cluster to cluster, these abundance patterns (apart from the MgAl anticorrelation that is not always found) 
have been observed in every well-studied GGC (Gratton, Sneden \& Carretta 2004). Despite  
the abundance of observational data, the theoretical interpretation of these empirical findings is not firmly established yet.
Internal nucleosynthesis and mixing can account for the C and N behavior in 
red giant branch (RGB) stars whose abundances 
exhibit a dependence on evolutionary status. These patterns, C destruction along with N production, are also found in field stars 
as they evolve up the red giant branch (RGB; Gratton et al. 2000). However, the identification of C, N, O, Na, Mg, and Al abundance variations in 
unevolved cluster stars (see, e.g., Briley et al. 1996; Gratton et al. 2001; Cohen \& Mel{\'e}ndez 2005) demonstrates that an external 
pollution mechanism must be the dominant source, since the interiors of unevolved stars are not hot enough to activate the necessary nucleosynthetic reactions (Langer, Hoffman \& Zaidins 1997). 
In addition, the evidence that the same amplitude of the anticorrelations are present in both 
unevolved and evolved stars of the same cluster -- whose convective envelope masses have very different values -- is an additional  hint that we are not  observing a simple surface pollution process, rather these stars were born from matter nuclearly processed by a previous stellar generation.  
Unlike C and N, abundance variations for the 
elements O, Ne, Na, Mg, Al are not present in field halo stars with the same [Fe/H] (Gratton et al. 2004 and references therein). 
This evidence represents a clear proof that the 
GGCs abundance patterns are associated with some (presently unknown) property of the cluster environment.

Intermediate-mass asymptotic giant branch (AGB) stars (see, e.g. Cottrell \& Da Costa 1981; Ventura \& D'Antona 2005) and massive stars 
(Prantzos \& Charbonnel 2006; Decressin et al. 2007) are candidates for the external pollution scenario; however, neither scenario currently 
provides a quantitatively satisfactory explanation. 

In these last few years, photometric observations have produced the unexpected result that also the CMDs of GGCs show clear signatures of 
multiple populations within the same cluster.  
NGC 2808 (D'Antona et al. 2005; Piotto et al. 2007), 
M54 (Siegel et al. 2007), NGC1851 (Milone et al. 2008) and NGC6388 (Piotto 2008) all show multiple stellar populations in their CMDs. 
The collected evidence is that multiple Main Sequences and/or Sub Giant Branches (SGBs) have been identified, and 
the most plausible explanation is the presence of stellar populations with distinct compositions and/or ages 
(see Cassisi et al. 2008 for a discussion on this issue).
In some cases like $\omega$ Cen (that has been for a long time the only example of GGC harboring multiple stellar populations) 
NGC 2808, the multiple Main Sequences can be interpreted as due to a large He-abundance 
variation (Norris 2004, D'Antona et al. 2005; Busso et al. 2007; Piotto et al. 2007 and references therein); for NGC1851 the population 
multiplicity 
can be attributed either to an age difference (Milone et al. 2008) or to a difference in the initial chemical heavy element pattern 
(Cassisi et al. 2008, Salaris, Cassisi \& Pietrinferni 2008). As recently discussed by Renzini (2008), the interpretation of these observations is lending 
increasing support to the idea that AGB stars are the most viable candidates for the pollution of the intra-cluster medium in the early 
stage of the GGC life.

So far, the empirical evidence that within all GGCs analyzed in detail groups of stars with different CNONa patterns do coexist, 
and that several clusters show also multiple populations in their CMDs, 
suggest a possible link between photometric and chemical population multiplicites. An illustrative case has been recently 
provided by NGC1851: its CMD displays a clear splitting in the SGB region, and at the same time Calamida et al. (2007) 
RGB Str\"omgren photometry (that is sensitive to the CN abundances) reveals the existence of two distinct branches. Starting from 
this evidence, Cassisi et al. (2008) and Salaris et al. (2008) have computed stellar models including 
for CNONa abundance anticorrelations with a (C+N+O) abundance increased by a factor of $\sim$2, that 
are able to reproduce the photometric SGB splitting without invoking any age difference 
between the two subpopulations. 
Very recently, Yong et al.~(2009, see also Yong \& Grundahl~2008) have demonstrated through accurate spectroscopical
 measurements that the [(C+N+O)/Fe] abundance ratio exhibits a range of 0.6 dex in this cluster. 
This empirical result supports the Cassisi et al.~(2008) scenario. In addition, Yong et al.~(2009) found that the 
abundances of Na, Al, Zr, and La abundances are correlated with the sum C+N+O, and therefore, NGC 1851 is the first cluster to provide strong 
support for the scenario in which AGB stars are responsible for the peculiar chemical patterns observed in GGC stars.
%

To assist in the interpretation of the photometric properties of GGC stars, it is therefore extremely useful to have available 
stellar models and isochrones calculated with metal mixtures reflecting some of these abundance pattern, different from a standard 
$\alpha$-enhanced mixture. 
Assuming that globular cluster formation and evolution processes are not significantly different from galaxy to galaxy, these 
models are important also to study the integrated colors and spectra of extragalactic unresolved globular clusters. The effect of these CNONa 
patterns and population multiplicity on the cluster integrated properties is a field until now not consistently and exhaustively investigated. 
A first contribution on this issues was made by Salaris et al. (2006), but only one single [Fe/H] value was accounted for; lately 
Cassisi et al. (2008) extended these early computations but once again only for one [Fe/H] value. In this work we make a further step forward, 
extending the calculations by Salaris et al. (2006) and Cassisi et al. (2008) to a wide [Fe/H] interval, that spans almost the entire range 
covered by the GGC system. All models will be made publicly available at the BaSTI official website (\url{http://www.oa-teramo.inaf.it/BASTI}).

The paper is organized as follows: \S~2 briefly summarizes 
the calculations, and comparisons with stellar models accounting for a normal $\alpha-$enhanced distribution are discussed in
\S~3. A summary and final remarks follow in \S~4.

\section{Inputs and models}

We have used the same stellar evolution code adopted for building the BaSTI library (see Paper I, II and III). 
The heavy-element mixture, with abundances representative of typical extreme values for 
the CNONa anticorrelations detected in GGCs -- denoted as 'extreme' mixture --  is the same as adopted 
by Salaris et al. (2006). It displays a 1.8 dex increase of the N abundance, a 0.6 dex decrease of C, a 0.8 dex increase of Na, and a 0.8 
dex decrease of O, compared to our adopted \lq{normal}\rq\ $\alpha-$enhanced heavy-element distribution (see Paper II) - 
we note that the scaled solar mixture adopted for the BaSTI library is taken from Grevesse \& Noels (1993).  
We did not include an anticorrelation between Mg and Al that is not always observed in GGCs  
because, as discussed in Salaris et al.~(2006) its effect on the models is negligible. 
The detailed heavy element distribution is reported in Table~1. 

We have taken into account 
the new heavy element distribution in both the radiative opacity and nuclear burning network. 
Details about the radiative opacities can be found in Salaris et al. (2006). Here we just mention that the low-temperature 
opacities for this new mixture have been computed consistently with the $\alpha$-enhanced tables by Ferguson et al.~(2005). 
In the comparisons that follow we will therefore 
consider the BaSTI $\alpha$-enhanced models computed using the Ferguson et al.~(2005) calculations at low temperatures.

The set of initial metal and He mass fractions $Z$ and $Y$ has been selected to obtain a grid of [Fe/H] values as close as 
possible (within at most 0.05 dex) to our $\alpha-$enhanced grid, because in individual GGCs the Fe abundance is found to be 
the same for both 'normal' $\alpha-$enhanced stars and stars showing CNONa abundance anticorrelations. 
An astrophysical meaningful comparison between the two 
sets of models should therefore be performed at constant [Fe/H]. In our evolutionary calculations we have tried to avoid, whenever possible, the 
interpolation in Z among the available radiative opacity tables, and this is the reason why the [Fe/H] values of the grid do not match 
exactly the corresponding values for the $\alpha-$enhanced grid. Differences with the corresponding grid points of the 
$\alpha-$enhanced model and isochrones are, as already noted, at most of only 0.05~dex, a difference that does not affect appreciably comparisons to evaluate the effect of the new mixture at fixed [Fe/H].
For a fixed [Fe/H], our mixture with CNONa anticorrelations displays a (C+N+O) sum that is a factor of $\sim$2 higher than in the 
$\alpha-$enhanced mixture. This is consistent, within the -- large -- observational error bars, with the spectroscopical estimates by Carretta 
et al. (2005) for the extreme values of the anticorrelations. 

At variance with the models presented in Papers I and II, we have now adopted the conductive opacity by Cassisi et al. (2007). 
As discussed by the authors, these new calculations affect only slightly the He-core mass at RGB tip and the luminosity of Horizontal Branch models, 
compared to the BaSTI calculations that employed the Pothekin (1999) conductive opacities. For the comparisons 
between results for this extreme mixture and our normal $\alpha-$enhanced models discussed in the rest of the paper,   
we have recomputed a selected set of $\alpha-$enhanced models using the conductive opacities by Cassisi et al. (2007).

To ensure a proper chemical composition coverage, we computed models spanning a large range 
of [Fe/H] values. The details of the chemical composition grid are given in Table~2.
For a fixed iron content  
we provide models for typically two He abundances. In case of [Fe/H]=$-0.69$ we account for four He contents, whereas for 
[Fe/H]=$-$0.87 and $-$0.56 one He content was taken into account. 
Since we wish to maintain fixed the iron content when varying $Y$, we have modified the reference value of $Z$ 
whenever differences in [Fe/H] induced by the change of $Y$ were larger than 0.02~dex.
For the whole [Fe/H] grid the lower $Y$ values (hereafter the 'standard' values of $Y$) have been obtained from $dY/dZ\sim 1.4$  and 
a cosmological $Y$=0.245, as in Papers I and II.
Notice that even the standard values of $Y$ are slightly larger that their counterparts in the $\alpha-$enhanced calculations with the 
same [Fe/H]. This is 
because a given [Fe/H] is obtained for $Z$ larger by a factor of $\sim$2 in case of the extreme mixture. Given that the $dY/dZ$ scaling is  
the same,  this explains the larger standard $Y$ values for the extreme mixture.
It is also important to notice that some enhancement of $Y$ is expected -- whichever is the source of external pollution -- to be present in 
GGC stars displaying the abundance anticorrelations.

We have computed stellar models in the mass range $\sim0.40M_\odot - 1.2M_\odot$, from the Zero Age Main Sequence (ZAMS) until the He-ignition 
at the RGB tip. We neglect the occurrence of core convective overshooting during the central H-burning stage. The less massive 
models ($M<0.7M_\odot$) with a MS lifetime longer than the Hubble  
time, have been calculated until central H exhaustion. We limited our computations to this mass range, given that these 
abundance anticorrelations 
are found in old stellar systems. For each chemical composition we have also computed an extended set of Horizontal Branch (HB)  
stellar models, by adopting the He-core mass and envelope He-abundance values obtained from a RGB progenitor 
whose age at the RGB tip is of the order of 13~Gyr. These HB models allow the computation of synthetic CMDs for HB 
populations with an arbitrary morphology (see, e.g., Salaris et al. 2008 for an application to study the HB of NGC1851). 
The RGB progenitors of these additional HB models have masses equal to $\sim 
0.8M_\odot$ at the lowest metallicities, increasing up to $\sim$1.0$M_\odot$ for the more metal-rich compositions. 

Each evolutionary track has been resampled to the
same number of points, to facilitate the computation of isochrones and 
their use in a population synthesis code. In brief, along each evolutionary
track some characteristic homologous points (key points -- KPs) 
corresponding to well-defined evolutionary phases have been identified. All tracks are then 
resampled considering the same number of points between two consecutive KPs. 
For a description of the adopted KPs we refer the reader to Paper~II. 
For all these new models, that extend from the ZAMS to the RGB tip, we take into account 8 KPs, i.e. the same number of 
KPs until the RGB tip as in all BaSTI models.
These resampled models have been then used to 
compute isochrones from 1.2~Gyr up to 16~Gyr, from the ZAMS up to the RGB tip.
Also the HB models have been resampled to the same number of points, using the same KPs as described in Papers~I and II..

Finally, all tracks (including HB tracks) and isochrones have been transformed to
various observational CMDs both in the Johnson-Cousins and in the ACS  
(Vega-mag) system, using the same color-$T_{eff}$ transformations and bolometric corrections adopted for the $\alpha-$enhanced 
models presented in Paper II and in Bedin et al.~(2005). These transformations are obtained from model atmospheres calculated 
with the $\alpha$-enhanced mixture of Paper II, and therefore are in principle 
not adequate for our extreme mixture. However, in broadband filters not bluer than approximately the Johnson $V$-band, the 
effect of this inconsistency should be minimized, since at least the color-$T_{eff}$ transformations are 
largely independent on the metals and their distribution (see Alonso et al.~1996, 1999, Cassisi et al.~2004).
As an additional, qualitative test to assess the effect of a different CNO abundance distribution (at constant [Fe/H]) on 
color transformations, we have made use of  
the recent theoretical spectra from the MARCS model atmosphere calculations 
(see Gustafsson et al. 2008)\footnote{Available at {\url http://marcs.astro.uu.se/}}. We have considered the available 
reference $\alpha$-enhanced mixture, that is very similar to our adopted one, 
and a `` Heavily CN-cycled composition'', whereby C is depleted and N is enhanced (by smaller amounts than 
in our extreme mixture) compared to the $\alpha$-enhanced one. This alternative mixture is very different from that adopted in our calculations, but  
gives at least a qualitative indication about the role played by the CN anticorrelation in determining color transformations and 
bolometric corrections. In the MARCS database there are no spectra with this composition for stars with gravities typical of main sequence stars.

We have compared bolometric corrections and selected colors at [Fe/H]=$-$1, for the following pairs of effective temperature 
(in Kelvin) and log($g$) ($g$ in cm \  s$^{-2}$ units) values: (4000, 0.0), (5000, 1.0), (6000, 2.5), considering the MARCS spherical models 
for a 1.0$M_{\odot}$ star, with microturbulent velocity equal to 2~km \ s$^{-1}$.
We found that colors like the Johnson-Cousins $(V-I)$ and near-IR colors are, as expected, completely unaffected by the 
CN anticorrelation included in the set of MARCS models. 
Changes of the order of 0.01-0.02~mag appear in the $(U-B)$ color at the two lowest temperatures. Bolometric corrections to the $U$ band are 
the most affected, at the level of at most $\sim 0.03$~mag.
Recent results from similar tests by Lee et al.~(2008) about the effect on broadband colors 
(determined from different calculations) of enhancing only C (or N) and keeping all other abundance fixed, are 
difficult to compare, even qualitatively, with ours. It is evident from their Table~3 that an enhancement of only C or N does alter 
especially the $(U-B)$ and $(B-V)$ colors of giants and dwarfs, but for our purposes it is relevant to consider a 
simultaneous enhancement of N and decrease of C.

We have also briefly extended our tests to Str\"omgren colors. 
All Str\"omgren colors involving the $u$ filter are heavily affected by the CN anticorrelation 
(e.g. $(u-y)$, $(u-b)$, $c_1$) at the level of $\sim$0.05~mag. 

On the basis of these qualitative results, a consistent (in terms of chemical composition). 
calculation of stellar model atmosphere and spectra is therefore necessary to produce adequate theoretical predictions for 
broadband filters in the blue/UV part of the spectrum, and to produce detailed high-resolution 
integrated spectra for studying extragalactic globular cluster systems. 

\section{The impact of a CNONa extreme metal distribution on stellar models and isochrones}

\noindent
In this section, we analyze briefly the properties of models and isochrones computed 
with our extreme CNONa composition, as compared to results with a standard $\alpha$-enhanced metal mixture.
We consider mainly the case \footnote{We note that similar conclusions are 
obtained when performing comparisons at a different iron content.}
of [Fe/H]=$-0.7$, at the metal rich end of 
the GGC [Fe/H] distribution, and compare the $\alpha-$enhanced models 
with $Z$=0.008, [M/H]=$-0.35$ and both Y=0.256 (from the BaSTI database) and Y=0.266, 
to the CNONa extreme models with $Z$=0.015, [M/H]=$-0.07$ and Y=0.266.
We remark that this comparison is performed by keeping the same iron content ([Fe/H]=$-0.7$) for both metal mixtures, that is 
the relevant case for GGCs, where [Fe/H] is essentially the same in stars 
with a normal $\alpha$-enhanced heavy element distribution and in stars showing CNONa anticorrelations. 
It is clear that a different 
result is obtained if we were to perform the same comparison at fixed global metallicity [M/H].

\begin{figure}        
\plotone{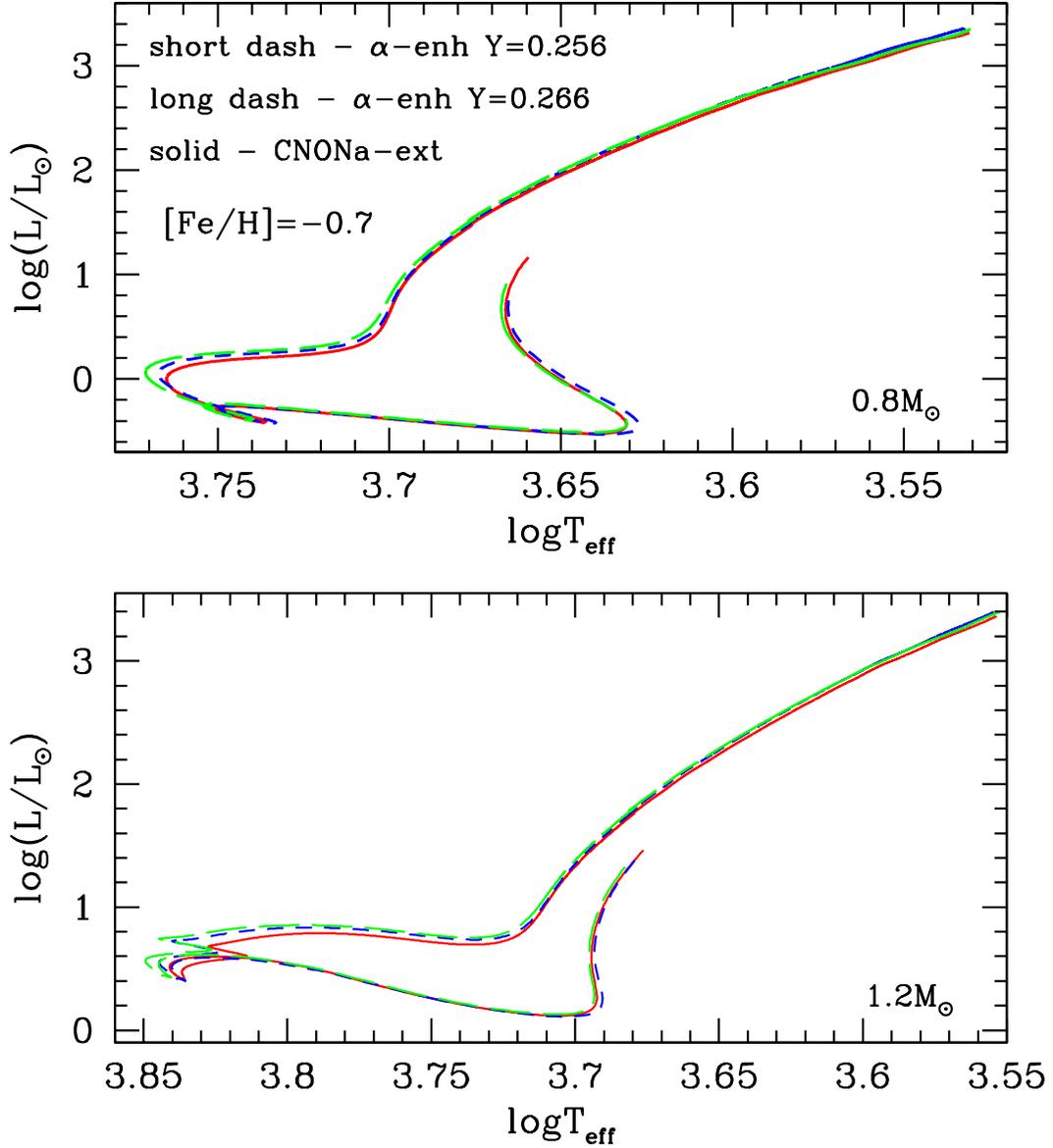}        
\caption{H-R location of models with mass equal to $0.8M_\odot$ and $1.2M_\odot$, computed with a CNONa extreme distribution (with $Y$=0.266) 
and an $\alpha-$enhanced one, but the same iron content [Fe/H]=$-$0.7. 
The $\alpha-$enhanced models have been computed for two different initial He contents ($Y$=0.256 as in the BaSTI database and $Y$=0.266). 
\label{trk1}}        
\end{figure}

Figure~\ref{trk1} displays the H-R diagrams of models with masses equal to $0.8M_\odot$ and $1.2M_\odot$. In 
case of the 0.8$M_\odot$ tracks, the MS overlaps almost perfectly. 
On the other hand, we can note that in case of the $1.2M_\odot$ model the MS is affected by the metal mixture. 
The Turn Off (TO) for the models with the extreme mixtures is in both cases fainter and cooler compared to the normal 
$\alpha$-enhanced calculations. The effect of changing $Y$ from 0.256 (as in the BaSTI database for 
the $\alpha$-enhanced mixture) to $Y$=0.266 in the $\alpha$-enhanced models amplifies the difference 
between the TO location of the extreme and normal models.
The TO evolutionary lifetimes of the two $0.8M_\odot$ tracks (at the same initial He abundance) are the same 
within $\sim1\%$ (the $\alpha-$enhanced model being slightly younger by about 120~Myr). 
In case of the $1.2M_\odot$ tracks, the $\alpha-$enhanced model is younger by 
about 71~Myr ($\sim3\%$) at the TO. 
This means that the extreme mixture affects -- as expected -- the evolutionary lifetimes of these low-mass stars 
because of the altered efficiency of the CNO burning (that in case of the $0.8M_\odot$ model gives 
only a marginal contribution to the energy budget) but overall the effect is small.

\begin{figure}        
\plotone{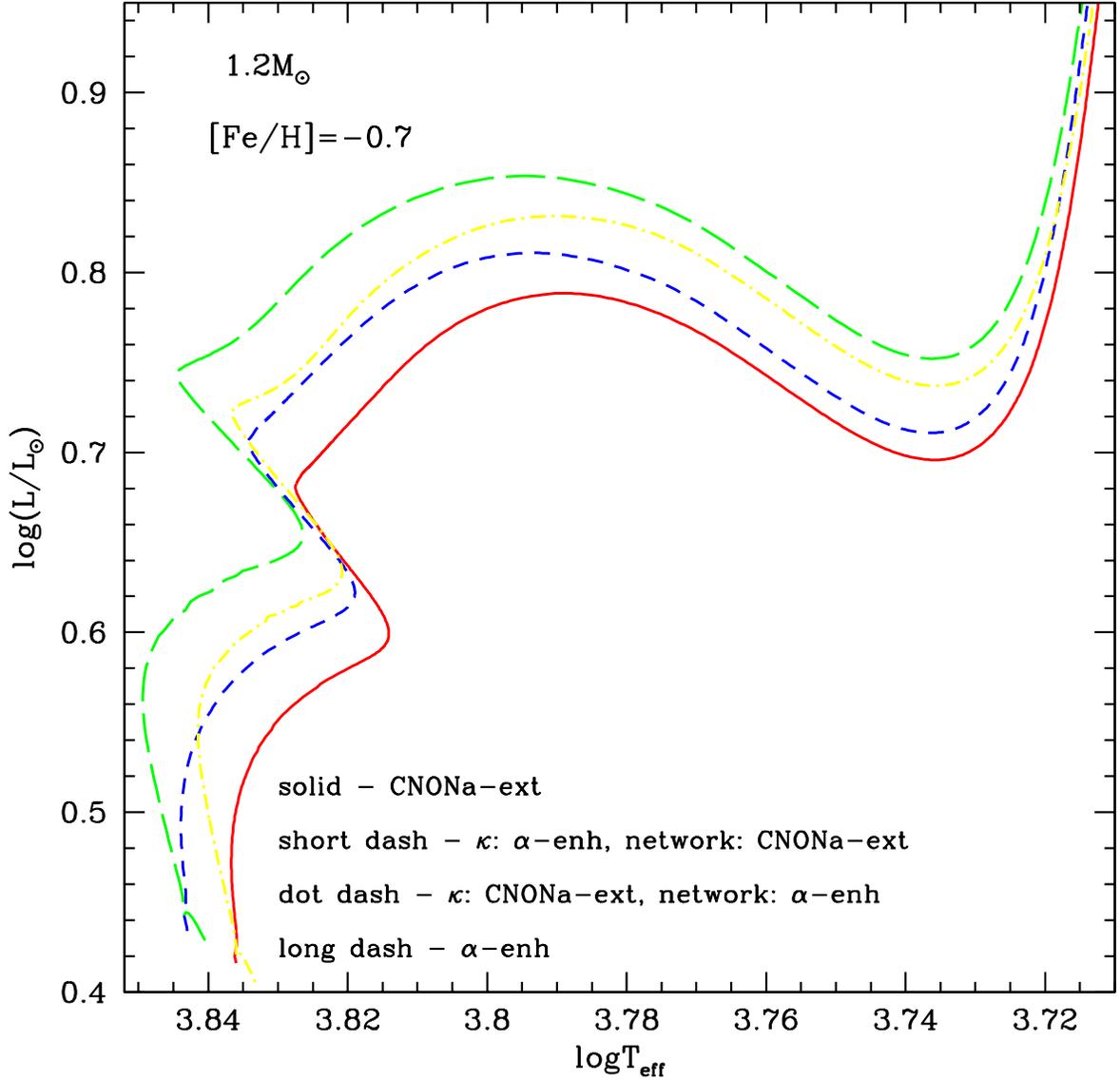}        
\caption{Comparison between $1.2M_\odot$ evolutionary tracks computed under different assumptions about the heavy elements mixture in 
the radiative opacities and nuclear burning network (see text for details). Reference evolutionary tracks of an $\alpha-$enhanced 
and a CNONa extreme model both with $Y$=0.266 and [Fe/H]=$-$0.7 are also displayed.  
\label{trk2}}        
\end{figure}

\begin{figure}
\plotone{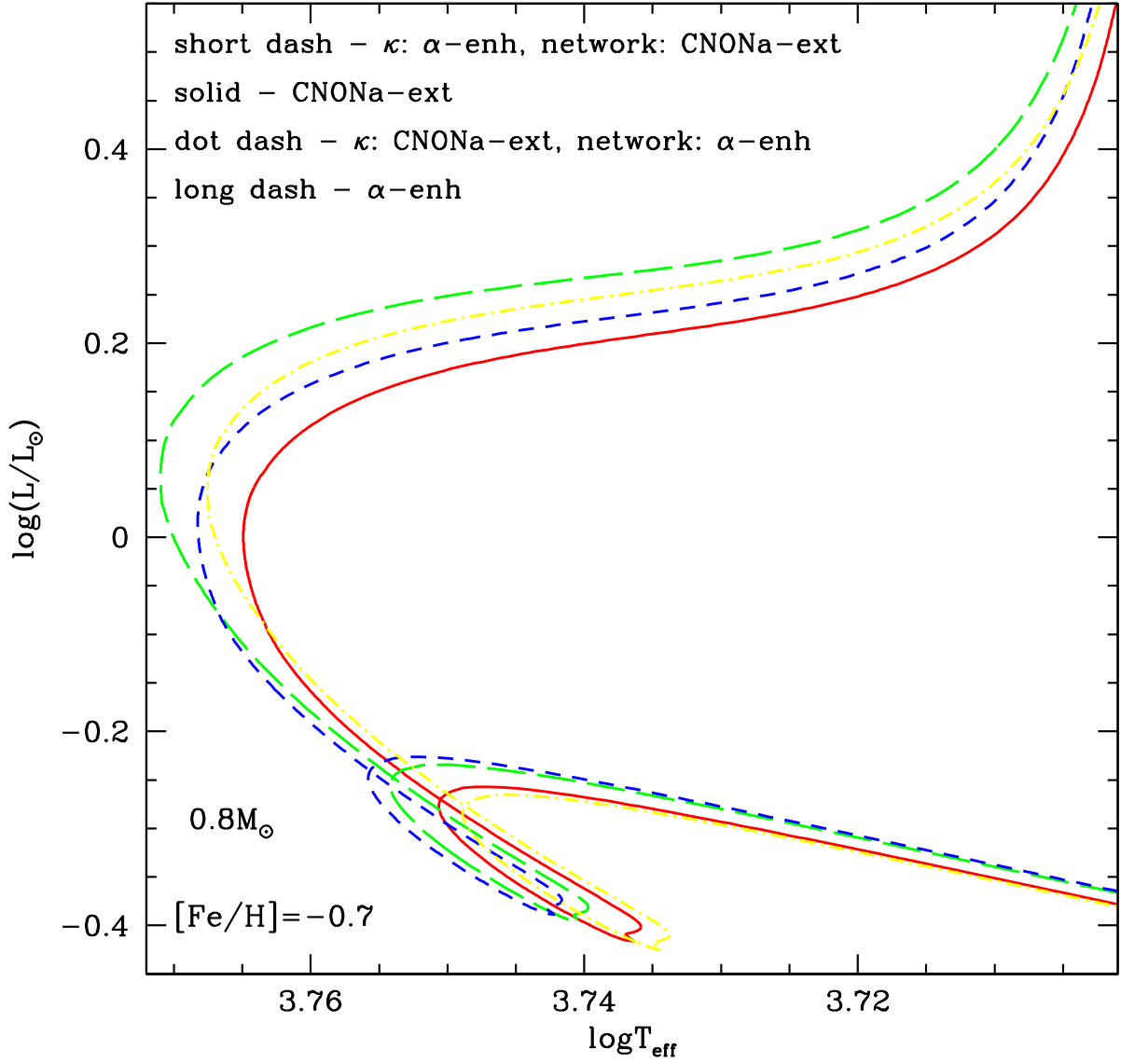}   
\caption{As in Fig. \ref{trk2}, but for $0.8M_\odot$ models.
\label{trk3}}   
\end{figure}

To disentangle the effects of changes in the opacity and in the nuclear network 
(changes in the 'nuclear network' denote changes in the efficiency of the burning processes 
caused by the variation of the CNO abundances between the two mixtures), due to 
the different metal mixtures, we computed additional models by varying
only one of these two inputs at a time. The results are shown in Figs.~\ref{trk2} and \ref{trk3} 
(all displayed tracks have been calculated assuming $Y$=0.266) and can be summarized as follows:

\begin{itemize}

\item 
When CNONa extreme models are computed employing the same chemical abundances of 
the $\alpha$-enhanced mixture in the nuclear network, the 
MS of the 0.8$M_\odot$ model is shifted to slightly cooler $T_{eff}$ by $\approx15~K$, whilst the MS of the 1.2$M_\odot$ model 
is displaced to generally higher 
$T_{eff}$ (the difference increasing as the core H-burning proceeds). In both cases the TO region 
is brighter and hotter.

\item
Using for the CNONa extreme model the opacities of the $\alpha$-enhanced one -- keeping everything else unchanged - moves the 
MS location toward higher $T_{eff}$ and the TO becomes also hotter and only slightly brighter, for both the 0.8 and 1.2$M_\odot$ models.

\item
In both numerical experiments the SGB becomes brighter and the base of the RGB only very slightly hotter by about $35~K$.
The largest effect on the SGB brightness is due to the change of nuclear network.

\end{itemize}

\begin{figure}        
\plotone{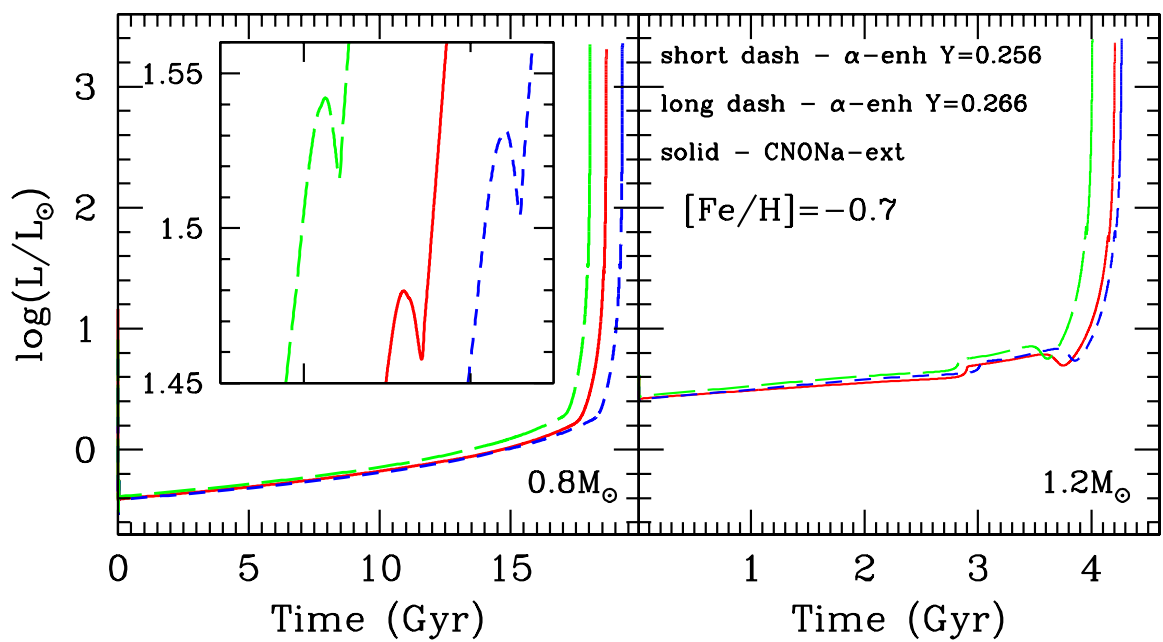}        
\caption{Run of the surface luminosity with time for the models shown in Fig.~\ref{trk1}. 
The inset in the left panel displays the behavior of the luminosity with time for $0.8M_\odot$ models when the H-burning shell encounters the chemical discontinuity left over by the first dredge up (the horizontal coordinates of the various tracks on this plot have been arbitrarily shifted, for the sake of clarity).
\label{trk4}}        
\end{figure}

Turning now the attention to the full RGB evolution, both sets of models in Fig.~\ref{trk1} converge to the same RGB locus.
Varying the initial $Y$ of the $\alpha$-enhanced models has a negligible effect, as expected, and overall the three displayed 
tracks show differences in temperature of at most only $\Delta{T_{eff}}\sim20~K$.
Figure~\ref{trk4} displays the surface luminosity as a function of time for the same stellar models in Fig.~\ref{trk1}. 
The inset enlarges the region around the RGB bump luminosity, i.e. when the H-burning shell encounters the 
chemical discontinuity in the envelope caused by the first dredge up during the early RGB phase. The 
CNONa extreme models have a RGB bump luminosity fainter by about $\Delta\log(L/L_\odot)\approx0.05$ compared to the reference 
$\alpha-$enhanced models with $Y$=0.256. The difference is slightly larger when $Y$=0.266 in adopted in the $\alpha$-enhanced model. 
This agrees with the result found by Salaris et al.~(2006) for [Fe/H]=$-$1.6.
It is worth noting that, despite the fact that the convective envelope reaches deeper layers in the CNONa extreme models, the amount of 
helium that is dredged up to the surface along the RGB is exactly the same ($\Delta{Y}=0.017$). 
This is due to the increased efficiency of the CNO cycle in the CNONa extreme models, that restricts 
the burning to more central regions. 

The comparisons in Figs.~\ref{trk1}, \ref{trk2}, \ref{trk3} and \ref{trk4} broadly confirm the results 
published over the years 
by Simoda \& Iben (1968, 1970), Renzini (1977), Salaris, Chieffi \& Straniero (1993), Rood (2000)\footnote{See also 
unpublished results specifically for a CNO-ehnanced mixture in  Rood, R.T., \& Crocker, D.A. 1997 
``Some Random Unpublished Results of Possible Interest,'', at  \url{http://www.astro.virginia.edu/~rtr/papers/}}, VandenBerg et al.~(2000), 
Dotter et al.~(2007), about the influence of individual chemical species on the evolution of low mass stellar models. 

The TO region (at fixed [Fe/H]) is completely controlled by the CNO abundance, through its effect on both opacity and nuclear network. The 
largest contribution appears to be due to the change of CNO in the nuclear network. On the other hand the MS location is 
mainly determined by the CNO abundance in the opacities.
The SGB luminosity is controlled again by the CNO abundance through its effect on both opacity and nuclear network. The RGB 
location is unaffected by the different metal mixture. Given that the $T_{eff}$ of the RGB controlled by the properties of the 
outermost layers (see, e.g., Salaris et al. 2002) this result confirms the expected result that the elements involved in 
the observed abundance anticorrelations (CNONa) do not contribute appreciably to the 
low temperature opacities, at least in the regime of the RGBs of GGCs.
Only the brightness of the RGB bump is changed, due to the increased total metallicity 
of the CNONa extreme mixture, i.e. through its effect on opacities in the temperature regime 
typical of the base of the convective envelope ($\approx 10^6$~K). 

We performed some additional numerical experiments by varying the degree of anticorrelation between CNONa. The general conclusion is that models with 
CNONa anticorrelations that preserve the CNO sum are indistinguishable from their $\alpha$-enhanced reference counterpart, at constant [Fe/H] (and $Y$). 
Differences in the TO and SGB region appear only when the CNO sum is not preserved. The RGB location is always unaffected. 

\begin{figure}
\plotone{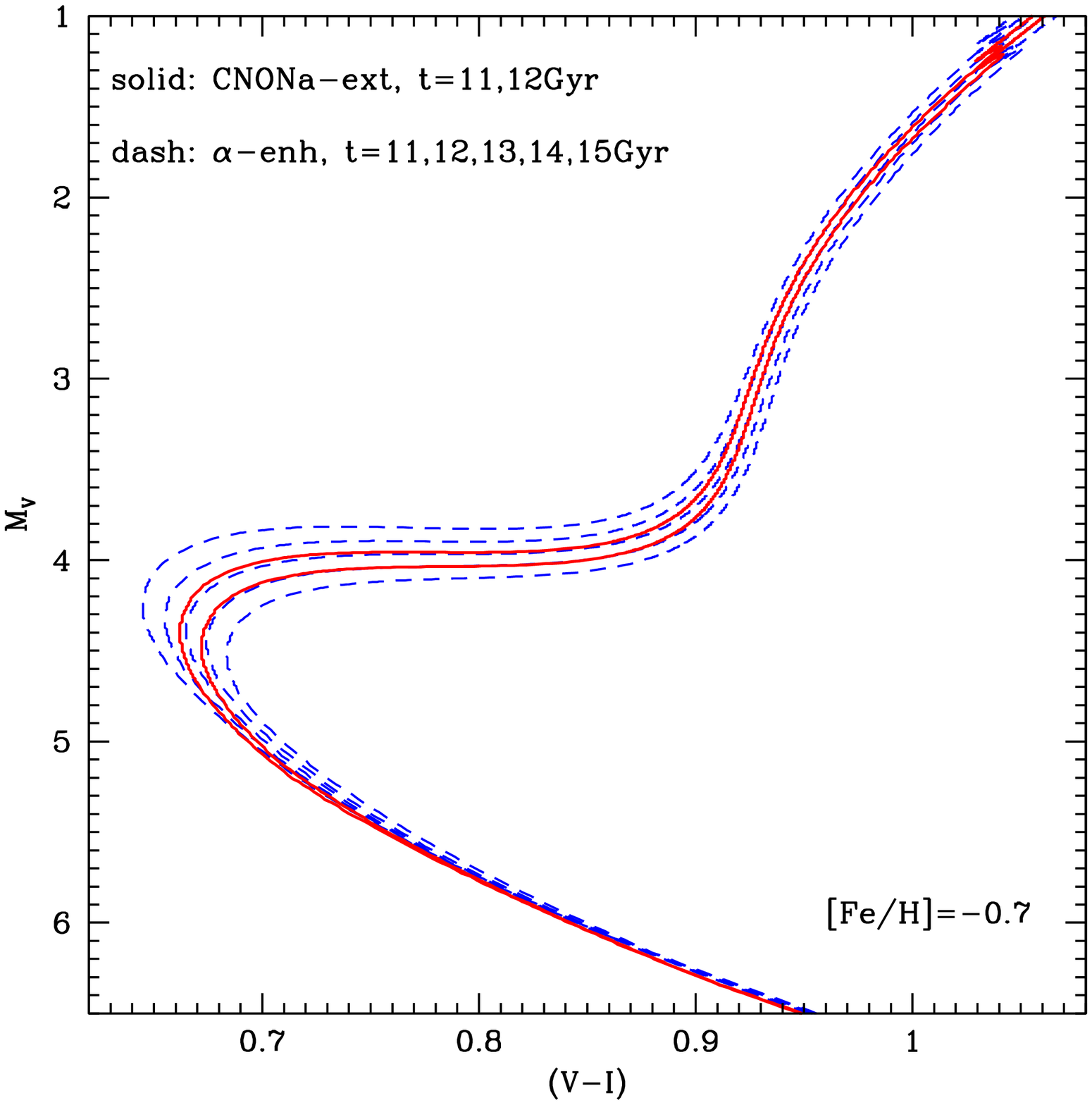}   
\caption{Comparison in the ($V, V-I$) CMD between selected [Fe/H]=$-$0.7 isochrones for the CNONa extreme mixture, and $\alpha$-enhanced 
isochrones (with $Y$=0.256) from the BaSTI archive.
\label{trk5}}   
\end{figure}

Figure~\ref{trk5} shows the ($V, V-I$) CMD of selected isochrones computed from CNONa extreme models, compared to 
our reference set of $\alpha-$enhanced isochrones retrieved from the BaSTI database (Pietrinferni et al. 2006, 
Manzato et al. 2008), with $Y$=0.256. As already noted by Cassisi et al. (2008) for a lower [Fe/H], the 
MS and RGB locations in this optical CMD are only very marginally affected by the use of a CNONa extreme mixture. It is clear that the same 
outcome is not expected to hold in case of bluer and/or intermediate- and narrow photometric bands such as the Str\"omgren photometric filters 
(see Calamida et al. 2007 and references therein). The two CNONa extreme isochrones for ages of 11 and 12~Gyr are almost perfectly  
matched at the TO and SGB regions by $\alpha-$enhanced isochrones with ages of 13 and 14~Gyr, respectively, i.e. there is 
a $\sim2$Gyr age offset between the TOs. However, one has to bear in mind that the CNONa extreme models have an initial He abundance larger by 0.01 by 
mass with respect the $\alpha-$enhanced, BASTI models. When accounting for this difference - see the previous discussion - the TO age 
offset is slightly decreased.

Moving on to the HB phase, one has to consider first the behavior of the He-core mass at the He-flash ($M_{cHe}$).
We emphasize once again that the models in the BaSTI archive have been 
computed by using the conductive opacity by Potekhin (1999), whereas the present CNONa extreme computations have been performed 
with the conductive opacity presented by Cassisi et al. (2007). To perform this differential comparison we have computed  
selected $\alpha-$enhanced models employing the same conductive opacities used for the CNONa extreme models.
The CNONa extreme 
$0.8M_\odot$ model has $M_{cHe}=0.4687M_\odot$ while the corresponding $\alpha-$enhanced model has $M_{cHe}=0.4753M_\odot$. This 
difference of the He-core mass at the RGB tip translates to $\Delta\log(L/L_\odot)\approx0.02$ for the RGB tip 
luminosity, and it is due to two effects. The first and dominant one is the 
heavy element distributions at fixed Fe abundance, implying different total metallicities $Z$. The second (minor) effect is 
related to the initial He abundance that in the BaSTI $\alpha-$enhanced model is $\Delta Y=$0.01 lower than the CNONa extreme one. 
To quantify this latter effect we have also computed a $0.8M_\odot$ track for the $\alpha-$enhanced mixture 
with Y=0.266. This model has $M_{cHe}=0.4734M_\odot$.

\begin{figure}
\plotone{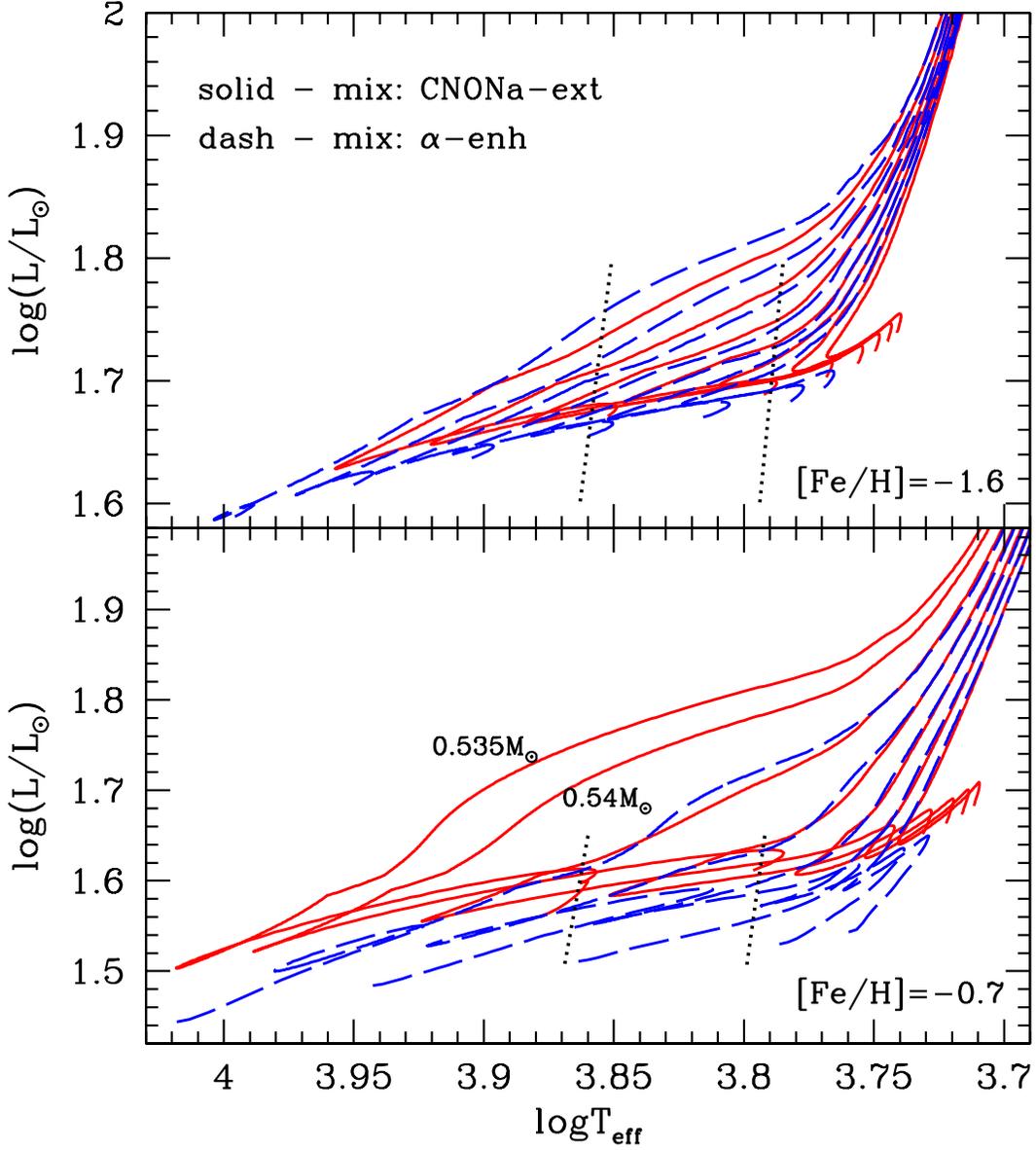}   
\caption{Bottom panel: HB models for [Fe/H]=$-$0.7 and the two adopted heavy element distributions. Except for the two models with labeled values, 
the mass of the displayed models is the same for both mixtures: it ranges between $0.55M_\odot$ and $0.59M_\odot$, 
in steps of $0.01M_\odot$. Top panel: as the bottom panel but for [Fe/H]=$-$1.6, 
in this case the mass ranges between $0.60M_\odot$ and $0.66M_\odot$, 
in steps of $0.01M_\odot$.
The approximate location of the red and blue edge of the RR Lyrae instability strip is also shown (dotted lines).
\label{trk6}}   
\end{figure}

Fig.\ref{trk6} displays the comparison between selected HB models for the two heavy element mixtures, and two values 
of [Fe/H]. 
Despite the lower value of $M_{cHe}$ and the larger $Z$, the CNONa extreme HB models are brighter than the 
reference $\alpha-$enhanced ones, because in the CNONa extreme models the H-burning through the CNO cycle is more efficient, 
as a consequence of the increased (C+N+O) sum. 
The brightness difference between the Zero Age HB (ZAHB) at $\log{T_{eff}}=3.83$  -- taken as representative of 
the mean effective temperature of the RR Lyrae instability strip -- is equal to $\Delta\log(L/L_\odot)\approx0.05$, i.e, $\approx0.12$ mag 
in the optical photometric bands, for models with [Fe/H]=$-$0.7. This value of $\Delta\log(L/L_\odot)$ decreases with decreasing metallicity, 
as can be appreciated by looking at the HB tracks for [Fe/H]=$-$1.6.

For a fixed total mass, the ZAHB location of the extreme models is cooler. However, as a 
consequence of the larger H-burning shell efficiency, the HB models for the extreme mixture perform 
more extended blue loop in the 
H-R diagram, as already shown by Salaris et al.~(2008). 
Due to the presence of these larger blue loops and the cooler ZAHB location, the mass range 
$\Delta M_{RR}$ spanned by the HB models able to cross the RR Lyrae instability strip is larger. 
For instance at [Fe/H]=$-$0.7 we find $\Delta M_{RR}\approx0.03M_\odot$, 
i.e. a factor of $\sim1.5$ larger than for compared to the reference $\alpha$-enhanced HB models 
($\Delta M_{RR}\approx0.03M_\odot$); while 
at [Fe/H]=$-$1.6 the same mass range is a factor $\sim1.3$ larger compared the $\alpha$-enhanced case.

\section{Summary and conclusions}

We have presented new evolutionary computations that extend the BaSTI stellar model library to account 
for a CNONa anticorrelation pattern (and varying initial He abundances) observed in GGCs, 
overimposed to the BaSTI reference $\alpha$-enhanced metal mixture. 

We confirm that the critical elements that determine the properties of isochrones with CNONa abundance variations 
are C, N and O. If the total CNO 
abundance is the same as in the reference $\alpha$-enhanced composition, one can safely use $\alpha$-enhanced 
isochrones to represent GGC subpopulations affected by CNONa anticorrelations. Conversely, in case of a different (C+N+O) sum 
-- and present spectroscopic determinations are consistent with a moderate variation of the total CNO abundance within stars in the same cluster --  
appropriate isochrones, like the ones presented in this paper, are necessary. In a CNO-enhanced mixture the isochrone TO 
(at fixed age, $Y$ and [Fe/H]) becomes dimmer and cooler, and the SGB fainter, compared to a reference metal distribution. 
The HB becomes brighter, the ZAHB location 
of HB tracks becomes cooler for a fixed total mass, whereas blue loops during the HB evolution get more extended, increasing the total mass range 
of objects that can potentially (depending on the RGB mass loss) populate the RR Lyrae instability strip. 

Linear interpolation between BaSTI $\alpha$-enhanced isochrones (and HB models) and these new calculations allows 
one to approximate the behavior of populations with varying degree of CNONa anticorrelations. 
Due to the method employed to resample all models with the same number of points between reference KPs, 
one needs simply to interpolate linearly between pairs of corresponding points along either isochrones or HB models.  
For these reasons our new calculations should represent a useful tool 
for investigating the overall stellar content and the occurrence of multipopulations in globular clusters.

\begin{figure}
\plotone{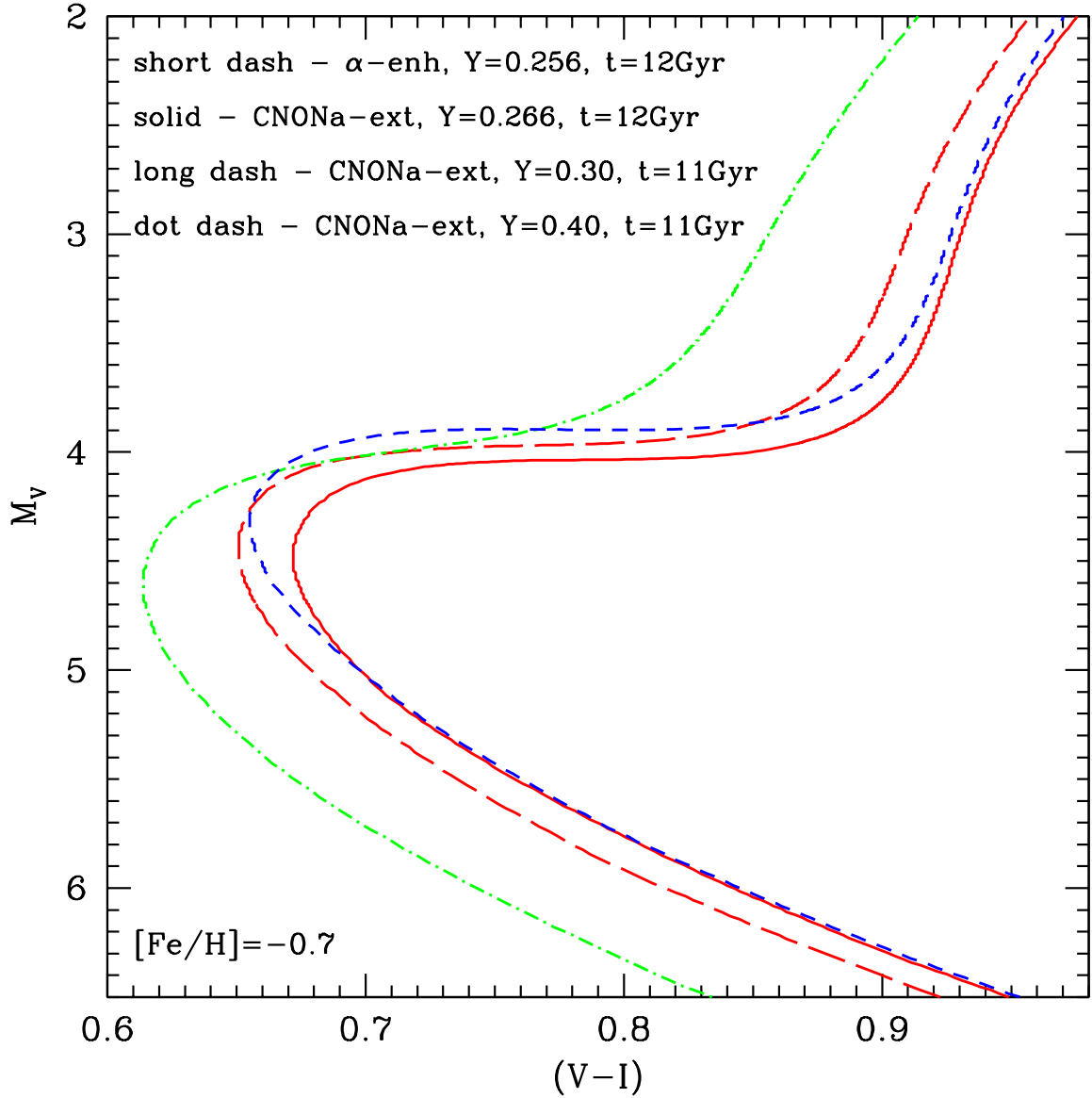}   
\caption{Comparison in the ($V, V-I$) CMD between selected [Fe/H]=$-$0.7 isochrones for the CNONa extreme mixture and various 
initial He contents (see labels), and an $\alpha$-enhanced isochrone (with $Y$=0.256) from the BaSTI archive.
\label{trk7}}   
\end{figure}

Before closing the paper, we show briefly a straightforward qualitative application of these models. The CMD of a cluster like 
NGC2808 (Piotto et al.~2007) displays a multiple main sequence, but single branches for the later evolutionary phases. 
On the other hand, a cluster like NGC1851 (Milone et al.~2008) and possibly NGC6388 (Piotto~2008) 
display a double SGB, but a single main sequence and RGB.
A substantial change in the initial He content between two distinct subpopulations is able to produce a main sequence and a SGB splitting, 
whereas a significant change in the CNO total abundance leaves unchanged the main sequence and affects TO and SGB. 
One can therefore ask the question of whether the multiple sequences in the CMD of NGC2808 and NGC1851 
are produced by different  combinations of age/$Y$/CNO variations between the cluster subpopulations.
Figure~\ref{trk7} compares various isochrones computed with selected choices of age, $Y$ and heavy element mixture. 
As expected on theoretical grounds, we assume that the stellar populations with larger initial He-content are also characterized by 
CNONa abundance anticorrelations. One can notice that the combination of a population with normal $\alpha$-enhanced metal abundances 
and a coeval one with CNONa anticorrelations and $Y$ very slightly enhanced, can reproduce qualitatively a CMD 
with just the SGB splitting. On the other hand, employing 
a subpopulation with CNONa abundance anticorrelations, ages younger by 1~Gyr and 
$Y$ increased by $\sim$0.05 causes a splitting along the main sequence and a SGB that is within $\sim0.1$~mag 
of the normal $\alpha$-enhanced one. Of course, detailed synthetic CMDs taking into account observational errors are 
needed to try to reproduce quantitatively multiple branches in the observed CMDs.

\acknowledgments
We wish to thank an anonymous referee for a very thorough report, comments 
and suggestions that have greatly improved the paper.
S.C. and A.P. acknowledge the financial support of INAF through the 
PRIN 2007 grant n. CRA 1.06.10.04: \lq{The local route to galaxy
formation}\rq, and PRIN MIUR 2007: \lq{Multiple stellar populations in globular clusters}.
This research has made use of NASA's Astrophysics Data System Abstract
Service and the SIMBAD database operated at CDS, Strasbourg, France.



\clearpage
\begin{deluxetable}{ccc}
\tablewidth{0pt}
\tablecaption{The adopted CNONa 'extreme' heavy element mixture}
\tablehead{
\colhead{element} &
\colhead{Number fraction}&
\colhead{Mass fraction} }   
\startdata
  C     &  0.013020    &    0.010454 \nl
  N     &  0.860012    &    0.805283 \nl
  O     &  0.054256    &    0.058031 \nl
  Ne    &  0.034240    &    0.046189 \nl
  Na    &  0.001970    &    0.003028 \nl
  Mg    &  0.013947    &    0.022661 \nl
  Al    &  0.000431    &    0.000777 \nl
  Si    &  0.010339    &    0.019412 \nl
  P     &  0.000041    &    0.000085 \nl
  S     &  0.005064    &    0.010853 \nl
  Cl    &  0.000046    &    0.000109 \nl
  Ar    &  0.000484    &    0.001292 \nl
  K     &  0.000019    &    0.000050 \nl
  Ca    &  0.001058    &    0.002836 \nl
  Ti    &  0.000066    &    0.000210 \nl
  Cr    &  0.000069    &    0.000241 \nl
  Mn    &  0.000036    &    0.000132 \nl
  Fe    &  0.004617    &    0.017238 \nl
  Ni    &  0.000285    &    0.001118 \nl

\enddata 
\label{mixture} 
\end{deluxetable}

\clearpage

\begin{deluxetable}{cccc}      
\tablewidth{0pt}      
\tablecaption{Initial chemical compositions of our model grid.}      
\tablehead{      
\colhead{[Fe/H]}&       
\colhead{[M/H]}&      
\colhead{$Z$}&       
\colhead{$Y$}}      
  \startdata   
      $-1.89$ &  $-1.27$  & 0.001  & 0.246   \\
                   &                &            & 0.280   \\
\hline
      $-1.60$ &  $-0.96$  & 0.002  & 0.248   \\
                   &                &            & 0.280   \\
\hline
      $-1.28$ &  $-0.66$  & 0.004  & 0.251   \\
                   &                &            & 0.280   \\
\hline
      $-0.97$ &  $-0.35$  & 0.008  & 0.256   \\
                   &                &            & 0.280   \\
\hline
      $-0.87$ &  $-0.25$  & 0.01  & 0.259    \\
\hline
      $-0.69$ &  $-0.07$  & 0.0150  & 0.266   \\
                   &                &  0.0143 &  0.300   \\
                   &                &  0.0133 &  0.350   \\
                   &                &  0.0123 &  0.400   \\
\hline
      $-0.56$ &  $0.06$  & 0.0198  & 0.2734    \\
\enddata      
\label{chemcomp}
\end{deluxetable}      



\clearpage
       
\clearpage 

\clearpage

\clearpage 
       
%
%

\clearpage

\clearpage

\clearpage

\end{document}